\documentclass[twocolumn,english,pra,showpacs,floatfix]{revtex4}
\usepackage[T1]{fontenc}
\usepackage[latin1]{inputenc}
\usepackage{graphicx}
\usepackage{amssymb}

\makeatletter

\usepackage{graphicx}

\usepackage{babel}
\makeatother

\newcommand \bea{\begin{eqnarray}}
\newcommand \eea{\end{eqnarray}}
\newcommand \ga{\raisebox{-.5ex}{$\stackrel{>}{\sim}$}}
\newcommand \la{\raisebox{-.5ex}{$\stackrel{<}{\sim}$}}
\newcommand{\av}[1]{\langle{#1}\rangle}

\begin{document}
\title{Hubbard model calculations of phase separation in optical lattices}
\author{H. Heiselberg}
\email{heiselberg@mil.dk}
\affiliation{Applied Research, DALO, Lautrupbjerg 1-5, DK-2750 Ballerup, Denmark}

\begin{abstract}
Antiferromagnetic, Mott insulator, d-wave and gossamer superfluid
phases are calculated for 2D square lattices from the extended Hubbard
(t-J-U) model using the Gutzwiller projection
method and renormalized mean field theory. Phase separation
between antiferromagnetic and d-wave superfluid phases is found near half filling
when the on-site repulsion exceeds $U\ga7.3t$, and coincides with a first
order transition in the double occupancy.
Phase separation is thus predicted for 2D optical lattices with ultracold Fermi
atoms whereas it is inhibited in cuprates by Coulomb frustration which 
instead may lead to stripes.
In a confined optical lattice the resulting
density distribution is discontinuous an with extended Mott plateau which
enhances the
antiferromagnetic phase but suppresses the superfluid phase.
Observation of Mott insulator, antiferromagnetic, stripe and 
superfluid phases in
density and momentum distributions and correlations is discussed.  
\end{abstract}
\maketitle


Ultra-cold atomic Fermi-gases present a new opportunity to study
strongly correlated quantum many-particle systems and to emulate high
temperature superconductors (HTc). Optical lattices realize the
Hubbard model, when the periodical lattice potential is strong enough
so that only the lowest energy band is populated, and
interactions, densities, temperatures,
etc., can be tuned.  Recent experiments with optical lattices have
measured momentum distributions and correlations, and have found
superfluid phases of Bose and Fermi atoms
\cite{Stoferle,Folling,Ketterle}, Mott insulators (MI)
\cite{Spielman,Schneider}, and band insulators \cite{Kohl,Rom}.

Long range Coulomb repulsion between electrons in cuprates prohibits
phase separation (PS), which in turn may lead to stripes, whereas PS
is allowed for atoms in optical lattices. The various competing phases
can be calculated in the 2D t-J-U model within the Gutzwiller projection
method and renormalized mean field theory (RMFT).
This method approximates the strong correlations and generally agrees well with
full variational Monte Carlo calculations \cite{Edegger,Laughlin}.
As will be shown below RMFT predicts PS at low doping
between an antiferromagnetic (AF) Neel order and a d-wave superfluid
(dSF) phase for sufficiently strong onsite repulsion $U\ga 7.3t$. The
amount of MI, AF and dSF phases, the density distribution and momentum
correlations in optical lattices are quite different from what would
be expected from HTc cuprates.

The t-J-U model was employed by Laughlin, Zhang and coworkers
\cite{Laughlin} to study AF, HTc and ``gossamer
superconductivity'' in cuprates and organic superconductors. Both the
Hubbard and t-J models are included in the t-J-U Hamiltonian
$H=H_U+H_t+H_s$ or
\bea \label{tJU}
 H   &=& U\sum_i \hat{n}_{i\uparrow} \hat{n}_{i\downarrow}
-t\sum_{\av{ij}\sigma} \hat{a}_{i\sigma}^\dagger \hat{a}_{j\sigma} \,+\,
  J\sum_{\av{ij}} {\bf S}_{i} {\bf S}_{j}\,,
\eea
where $\hat{a}_{i\sigma}$ is the usual Fermi 
creation operator, $\sigma=(\uparrow,\downarrow$) is the two hyperfine states 
(e.g. ($-\frac{9}{2},-\frac{7}{2}$) for $^{40}$Na), 
$n_{i\sigma}=\hat{a}_{i\sigma}^\dagger \hat{a}_{i\sigma}$ the density, 
${\bf S}_i=\sum_{\sigma\sigma'}\hat{a}_{i\sigma}^\dagger\vec{\sigma}_{\sigma\sigma'}
\hat{a}_{i\sigma'}$
and $\av{ij}$ denotes nearest neighbours.  $U$ is the on-site 
repulsive interaction, $t$ the nearest-neighbour hopping
parameter and $J$ the spin-spin or super-exchange coupling. 

The t-J-U model allows for doubly occupied sites and thereby also MI
transitions. As both are observed in optical lattices the t-J-U model
is more useful as opposed to the t-J model which allows neither.  For large $U/t$ the
t-J-U and Hubbard models reduce to the t-J model with spin-spin
coupling $J=4t^2/U$ due to virtual hopping. At finite $J$ and $U$ the
t-J-U model is to some extent double counting with respect to the
Hubbard model with $J=0$.  However, when RMFT is applied the virtual
hopping is suppressed in the Hubbard model which justifies the explicit
inclusion of the spin Hamiltonian as done in the t-J-U model.
Also, optical ``superlattices'' provide
additional spin-spin interaction 
\cite{Klein} and thus realize the unconstrained t-J-U model.

The t-J-U model on a 2D square lattice at zero temperature was studied
for various onsite coupling and densities in
Refs. \cite{Laughlin,Yuan} but mainly for $J\simeq 0.3-0.5t$, which is
the relevant case for cuprates and organic superconductors.  In
optical lattices the Hubbard model can be realized with any on-site
repulsion by tuning near a Feshbach resonance.  
For strong onsite repulsion spin-exchange requires $J=4t^2/U$ and
we will therefore study the t-J-U model with this parameter
constraint. However, keeping in mind that for more moderate $U$ double counting
may occur leading to an effective value for $J$. 
We include nearest-neighbour hopping and interactions
only and therefore the phase diagram is symmetric around half filling
$n=1$, where $n=N/M$ is the density or filling fraction ($N$ is the
number of electrons or atoms in $M$ lattices sites).

In the Gutzwiller approximation the trial wave function
$|\psi\rangle=\Pi_i[1-(1-g)\hat{n}_{i\uparrow} \hat{n}_{i\downarrow}]
|\psi_0\rangle$ 
is a projection of the Hartree-Fock wave function $\psi_0$. The
variational parameter $g$ suppresses double occupancy and 
varies between 0 corresponding to no doubly
occupied sites ($U\to\infty$) and 1 corresponding to no correlations  $(U=0)$.
In the Gutzwiller 
projection method the spatial correlations are included only
through the renormalization factors $g_t$ and $g_s$
defined such that $\langle H_s\rangle=g_s\langle H_s\rangle_0$ 
and $\langle H_t\rangle=g_t\langle H_t\rangle_0$, where the expectation values
are with respect to $\psi$ and $\psi_0$ respectively.
The renormalization factors are calculated by classical statistics
\cite{Ogawa,Vollhardt}
\bea
  g_s = \left( \frac{n-2d}{n-2n_+n_-} \right)^2 \,,
\eea
for the spin term and for the hopping term
\bea \label{hop}
  g_t &=& \sqrt{g_s}
  \left[\sqrt{\frac{x_-}{x_+}(x+d)}+\sqrt{\frac{n_-}{n_+}d}\right] \nonumber\\
  &&\left[\sqrt{\frac{x_+}{x_-}(x+d)}+\sqrt{\frac{n_+}{n_-}d} \right] \,.
\eea
Here $d$ is the double occupancy,
$n_\pm=n/2\pm m$ the mean up and down occupation numbers
for a magnetization $m$ at each site, $x=1-n$ the doping, and $x_\pm=1-n_\pm$.
The double occupancy projection factor is given by
$g^2=d(x+d)/(x_+x_-n_+n_-g_s)$.

The resulting energy is by Gutzwiller projection
\bea
   E = MUd + g_t\langle H_t\rangle_0 + g_s\langle H_s\rangle_0 \,.
\eea

The RMFT equations for the t-J-U model in the Gutzwiller projection method
have been derived \cite{Yuan}
in terms of the order parameters for d-wave pairing, hopping average
and staggered magnetization (with commensurate nesting vector
${\bf q}=(\pi,\pi)$, in units where the lattice constant is unity)
defined as
\bea \label{Dorder}
   \Delta_{\av{ij}}&=&\langle \hat{a}_{i\downarrow}\hat{a}_{j\uparrow} -
\hat{a}_{i\uparrow}\hat{a}_{j\downarrow}\rangle_0 =\pm\Delta \,,\\
   \chi &=& \langle \hat{a}_{i\uparrow}^\dagger\hat{a}_{j\downarrow} +
\hat{a}_{i\downarrow}^\dagger\hat{a}_{j\uparrow}\rangle_0  \,, \\
   m&=& (-1)^i\langle \hat{a}_{i\uparrow}^\dagger\hat{a}_{i\downarrow} +
\hat{a}_{i\downarrow}^\dagger\hat{a}_{i\uparrow}\rangle_0 /2 \,,
\eea
respectively. The $(+/-)$ in (\ref{Dorder}) corresponds to
a difference between neighbour sites  $\av{ij}$
of one lattice unit in the $(x/y)$ direction respectively.

The resulting variational energy is \cite{Yuan}
\bea
  E/M &=& Ud-\tilde{\mu} x -\frac{1}{2M}\sum_k 
        \left(E_{k}^++E_{k}^-\right) \nonumber\\
 &&+\frac{3}{4}g_sJ(\Delta^2+\chi^2)+2g_sJm^2 \,.
\eea
Here, the AF and dSF couples four band energies 
$E_{k}^\pm=\sqrt{(\xi_k\mp\tilde{\mu})^2+(\Delta_d\eta_k)^2}$
and $-E_{k}^\pm$, with
$\xi_{k}=\sqrt{\epsilon_k^2+\Delta_{af}^2}$,
$\epsilon_k=-(g_tt+3g_sJ\chi/8)\gamma_k$,
$\gamma_k=2(\cos k_x+\cos k_y)$ and $\eta_k=2(\cos k_x-\cos k_y)$.
$\Delta_d=3g_sJ\Delta/8$ and $\Delta_{af}=2g_sJm$
are the dSF and AF order parameters.
The Lagrange multiplier $\tilde{\mu}$ differs from the chemical potential
because the renormalization factors depend on density as will be discussed
later.

Varying the free energy $F(\Delta,\chi,m,d,\tilde{\mu})=E-\tilde{\mu}N$ 
with respect to its
five parameters leads to the ``gap'' equations for the pairing
gap, hopping average and density
\bea \label{gap}
 \Delta &=&\frac{1}{8M} \sum_k \eta_k^2\Delta_d 
      \left( \frac{1}{E_{k}^+}+ \frac{1}{E_{k}^-}\right)\,, \\
 \chi &=& -\frac{1}{8M} \sum_k \gamma_k\frac{\epsilon_k}{\xi_k} 
 \left(\frac{\xi_k-\tilde{\mu}}{E_{k}^+}+\frac{\xi_k+\tilde{\mu}}{E_{k}^-}\right)\,, \\
 x &=& \frac{1}{2M} \sum_k  \left( \frac{\xi_k-\tilde{\mu}}
 {E_{k}^+}-\frac{\xi_k+\tilde{\mu}}{E_{k}^-}\right)\,. 
\eea
The gap equations that determines the magnetization $m$ and double
occupancy $d$ \cite{Yuan} are more complicated since the
renormalization factors $g_s$ and $g_t$ also depend on $m$ and $d$.

\begin{figure}
\includegraphics[scale=0.42,angle=-90]{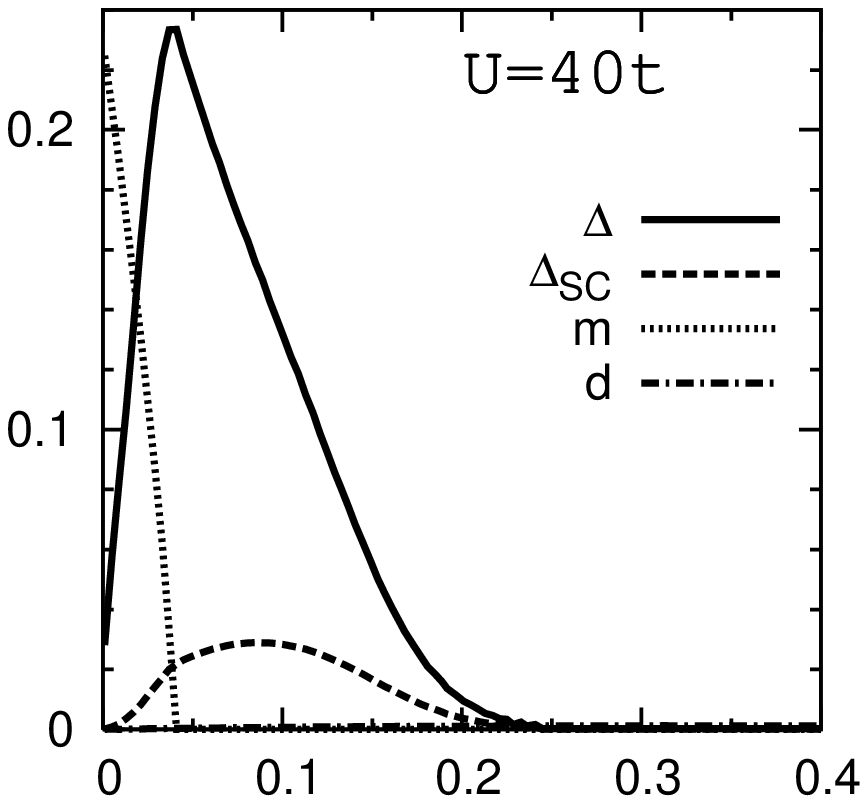}
\includegraphics[scale=0.42,angle=-90]{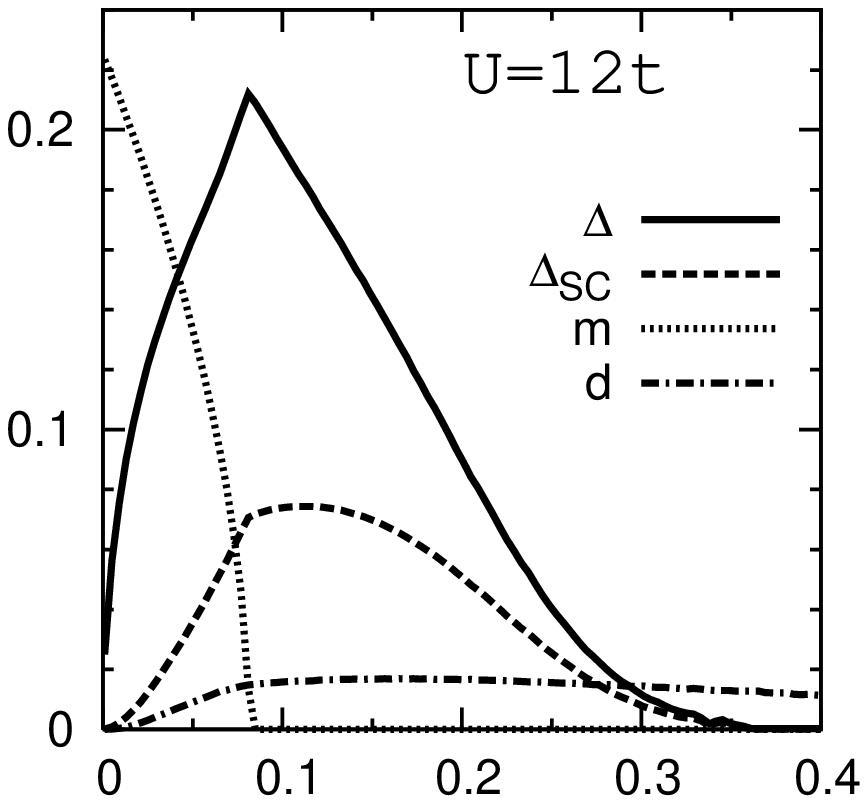}
\includegraphics[scale=0.42,angle=-90]{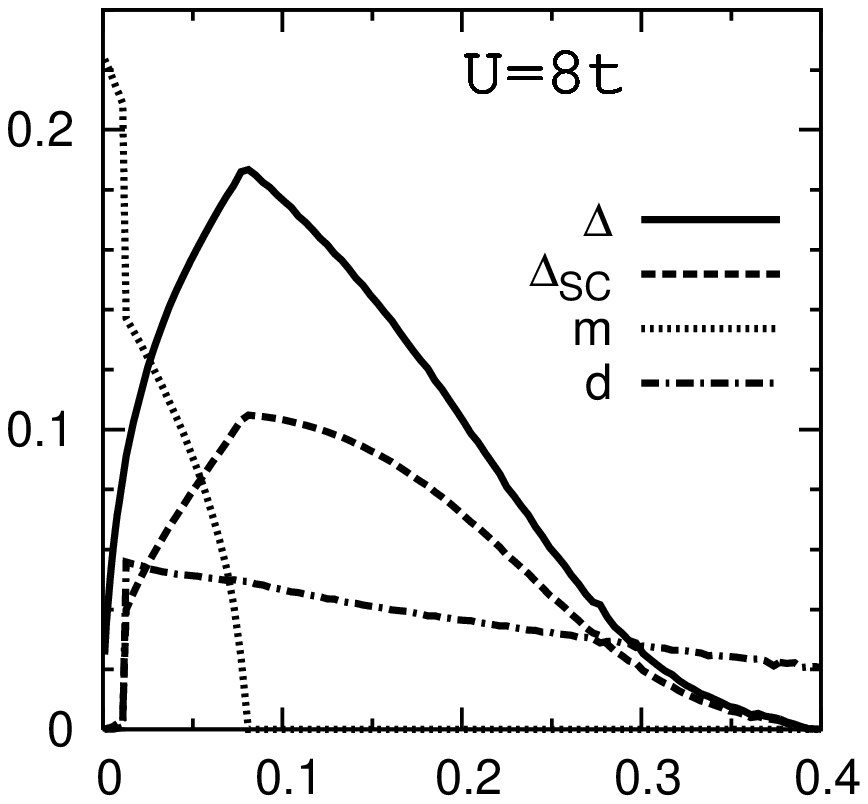}
\includegraphics[scale=0.42,angle=-90]{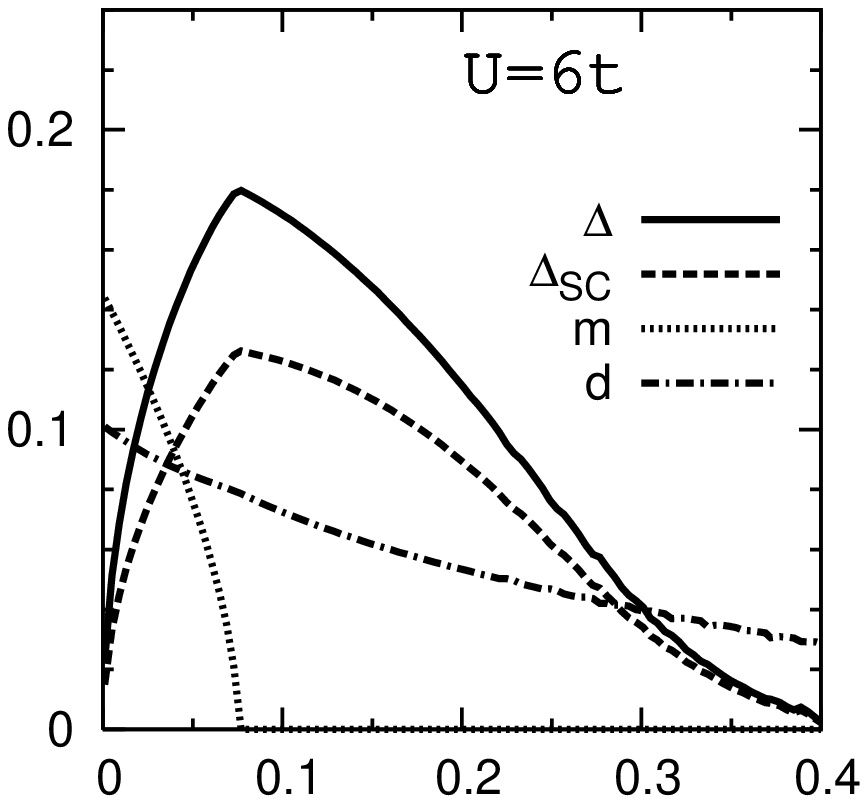}
\caption{Order parameters vs. doping $x=1-n$ for 
$J/t=4t/U=0.1,1/3,0.5,2/3$ in RMFT for the 2D t-J-U model. 
\label{fig1}}
\end{figure}

The resulting order parameters for the t-J-U model constrained to the
Hubbard or t-J model limit $J=4t^2/U$ are presented in Fig. 1 for
$U/t=$40, 12, 8 and 6 corresponding to $J/t=$ 0.1, 1/3, 0.5 and 2/3
respectively.
At half filling (x=0) we find a first order phase transition in which
the double occupancy jumps from zero to a finite value for $J\ge
0.55t$ ($U\le 7.3t$) similar to Refs. \cite{Laughlin,Yuan}. 
The ground state remains an AF for smaller U
without any dSF. This is in agreement with expectation for the 2D 
Hubbard model but in contrast to the results for a fixed
$J=t/3$ \cite{Yuan}, where the AF vanishes for $U\ge 7t$ and a dSF
appears for $U\ge 5.3t$ such that a coexisting AF and dSF at half filling
(a ``gossamer superfluid'') exists between $5.3\le U/t\le 7$. 
The reason for this difference traces back to the larger
value for $J=4t^2/U$ in our calculation which stabilizes the AF phase
and destabilizes dSF because the corresponding two order parameters
compete.  The first order transition in the double occupancy is,
however, robust in the sense that it is driven by the on-site
repulsion and remains at a critical value $U\sim 7-9t$ even when $J$ is
allowed to vary independently from the $J=4t^2/U$ constraint.
The case $J=0.5t$ ($U=8t$) is close to the critical coupling and we find a
transition in $d$ at low doping as shown in Fig. 1 that also affects the
other order parameters $m$ and $\Delta_{SC}$.

The gap depends sensitively on coupling, magnetization
and double occupancy. If $m=d=0$ the gap equation gives
$\Delta=\chi=\sum_k\sqrt{\cos^2k_x+\cos^2k_y}/(2\sqrt{2}M)\simeq 0.339$ 
at half filling.
At low densities the gap equation simplifies because there is no AF
order $\Delta_{af}=0$. The gap equation depends on the pairing to
quasi-particle energy ratio $2\Delta_d/\epsilon_{k=0}= V\Delta$,
where $V\equiv 1/(\chi+8g_tt/3g_sJ)\simeq 3J/4t$ is the effective pairing
coupling divided by the bandwidth.
Also the Fermi surface becomes
circular with Fermi momentum $k_F=\sqrt{2\pi n}$ and
$\tilde{\mu}=(k_F^2-4)(g_t+3g_sJ\chi/8)$.
The r.h.s. of the gap equation (\ref{gap})
can then be calculated analytically to leading logarithmic order 
in the effective coupling and to leading orders in the density.
The resulting gap becomes
\bea
   \Delta =  \frac{1}{V\sqrt{n}}
   \exp\left[-\frac{4}{\pi n^2}(\frac{1}{V}-c_0-c_1n-c_2n^2)\right] \,.
\eea
The higher order corrections in density 
can be calculated numerically:
$c_0\simeq0.27$, $c_1\simeq0.57$ and $c_2\simeq0.09$. 

The AF order parameter is defined as the expectation value as in
Eq. (\ref{Dorder}) but for the w.f.  $|\psi\rangle$ in stead of
$|\psi\rangle_0$.  It therefore also attains a renormalization factor
$m_{AF}=\sqrt{g_s}m$ \cite{Yuan}.
Likewise the dSF order parameter is renormalized $\Delta_{SC}=g_\Delta\Delta$ with 
$g_\Delta=\frac{\sqrt{g_s}}{2}
  ([\sqrt{\frac{x_-}{x_+}(x+d)}+\sqrt{\frac{n_-}{n_+}d}]^2
+[\sqrt{\frac{x_+}{x_-}(x+d)}+\sqrt{\frac{n_+}{n_-}d} ]^2)$,
and is shown in Fig. 1. We expect that the dSF critical
temperature is similar.

\begin{figure}
\includegraphics[scale=0.65,angle=-90]{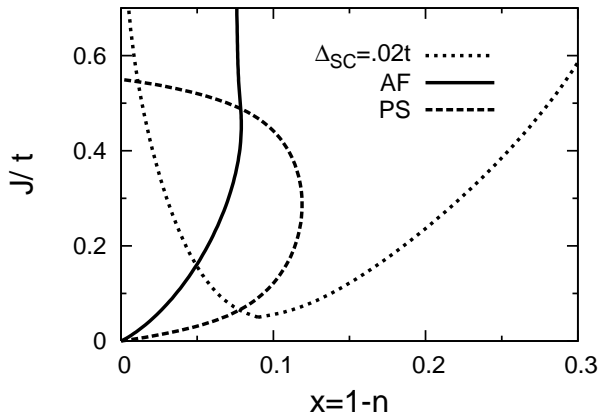}
\caption{Phase diagram for the t-J-U model under the constraint $J=4t^2/U$.
AF and PS occur for $|x|$ below respective curves. Above dotted curve the
dSF gap exceeds $\Delta_{SC}\ge 0.2t$. 
At half filling the first order transition to double occupancy 
above $J\ge0.55t$ is complementary to PS (see text).  \label{fig2}}
\end{figure}

Since the renormalization factors depend on density the chemical potential 
$\mu = dE/dN$ differs from the Lagrange multiplier $\tilde{\mu}$ by
\bea
  \mu  = \tilde{\mu} + \av{H_t}_0\frac{\partial g_t}{\partial n}+
        \av{H_s}_0 \frac{\partial g_s}{\partial n} \,.
\eea
Near half filling the chemical potential decreases with increasing
density in the AF phase i.e. the energy dependence on density is
concave. This is unphysical and signals PS to an AF phase at half
filling coexisting with a dSC phase at a density $|x|\la 0.14$ shown
in Fig. 2 that is determined by the Maxwell construction.  The PS is
found to terminate at coupling $J\simeq0.55t$ ($U\simeq7.3t$) where
the double occupancy undergoes a first order transion from zero to a
finite value. Relaxing the $J=4t^2/U$ constraint does not change the phases
or PS qualitatively except for gossamer dSF.

 Whereas PS is prohibited in cuprates by long range Coulomb repulsion between
electrons, PS is permitted for the neutral atoms trapped in
optical lattices where it leads to discontinuities in 
the density distribution $n(r)$ vs. trap radius $r$. 
For a large number of trapped atoms in a shallow
confining potential, the Thomas-Fermi approximation applies (see,
e.g. \cite{HH}). The total chemical potential is then given by the
local chemical potential $\mu(n)=dE/dN$ and the trap potential, which
is on the form $V_2r^2$ in most experiments,
\bea \label{mu}
 \mu(n) +V_2r^2 = \mu(n=0)+V_2R^2 \,.
\eea
The sum must be constant over the lattice and can therefore be set to
its value at the edge or radius $R$ of occupied lattice sites, which
gives the r.h.s. in (\ref{mu}). The chemical potential for the dilute
lattice gas in the 2D Hubbard model is $\mu(n=0)=-4t$ whereas just
below half filling it lies between $\mu(n=1)=0$ when $U=0$ and
$\mu(n=1)=4t$ when $U=\infty$. Correspondingly the cloud size, when PS
or a MI phase appears in the centre, lies between $R_c\le R\le 2R_c$,
where $R_c=2\sqrt{t/V_2}$, and the number of trapped particles
$N=2\pi\int^R_0 n(r)rdr$ is of order $N_c=\pi R_c^2$.  When more
particles are added a MI plateau forms in the centre (see Fig. 3)
because the chemical potential has a gap $\Delta\mu$ at half filling.
This gap and the chemical potential above half filling can be found
from particle-hole symmetry which implies $E(-x)=E(x)+MUx$, such that
$\mu(-x)=U-\mu(x)$.  We find that as $U$ decreases so does $\Delta\mu$
but it remains finite even though the double occupancy undergoes a
first order transition to a finite value, i.e. the MI remains along
with the AF phase.  The approximation $J=4t^2/U$ is, however, only a
valid for large $U$ and the constrained t-J-U model may not describe
the Hubbard model well for large $J$.  The MI plateau remains in the
trap centre until $N\ga N_c\Delta\mu/t$ corresponding to a trap size
$R\ga\sqrt{\Delta\mu/V_2}$. Increasing the number of trapped particles
further enforces doubly occupied sites and eventually a band insulator
with $n=2$ in the centre as seen in Fig. 3.

\begin{figure}
\includegraphics[scale=0.6,angle=-90]{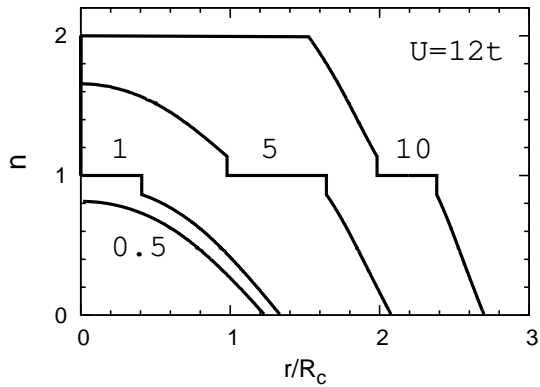}
\caption{Density distributions of Fermi atoms in an optical lattice
confined by a harmonic trap for $U=12t$ ($J=t/3$) filled with the number
of particles $N/N_c=$ 0.5, 1, 5, and 10. The density discontinuities from 
$n\simeq 1.14$ and $n\simeq0.86$ to the MI at $n=1$ due to PS between 
dSF and AF are absent for stripe phases (see text).}
\end{figure}

The density distributions and MI plateaus have been measured
experimentally for Bose atoms in optical lattices by, e.g.,
differentiating between singly and doubly occupied sites
\cite{Folling}. A similar technique applied to Fermi atoms might
observe the transition in $d$ at $U\simeq7.3t$. Recently Schneider et
al. \cite{Schneider} have measured column densities for Fermi atoms in
optical lattices and find evidence for incompressible Mott and band
insulator phases. Even lower temperatures are required for
observing the AF and dSF phases and the density discontinuities due to PS.

Spin and charge density waves in form of stripes are not included in
the above RMFT calculations. Stripes are observed in several cuprates
\cite{Tranquada} whereas numerical calculations are model dependent. 
Long range Coulomb frustration can explain why PS is
replaced by stripes and an AF phase at very low doping \cite{HH} as
observed in cuprates. Cold atoms in optical lattices can discriminate
between the PS and stripe phases since the latter does not have
density discontinuities.

Bragg peaks have been observed for bosons in 3D \cite{Folling} and 2D
\cite{Spielman} optical lattices, and dips for 3D fermions in
\cite{Rom}.  The Bragg peaks and dips occur in momentum correlation
functions $C({\bf k},{\bf k}')$ when the relative momentum is ${\bf q}\equiv{\bf
  k}-{\bf k}'=\pi(n_x,n_y)$, where $n_x,n_y$ are even integers
\cite{Bloch}. In an
AF phase the periodicity of a given spin is two lattice distances and
dips also appears for odd integers. 
Just as for the AF phase we can in a stripe phase expect
charge and spin correlations as in low energy magnetic neutron
scattering \cite{Tranquada} 
at incommensurate ${\bf q}=2\pi(0,\pm 2x)$ and
${\bf q}=\pi(1,1\pm 2x)$ respectively, observed as
dips at these wave-numbers.  However, because the doping $x$
varies in the trap, the Bragg dips are distributed
over the range of values for $x$ and are therefore harder to
distinguish from the background. If, however, the four stripe
periodicity occurs for $1/8\le x\le 1/4$ as in cuprates,
we can expect novel Bragg dips for charge correlations at ${\bf
  q}=(0,\pm\pi/2)$ and ${\bf q}=(\pm\pi/2,0)$ and for spin
correlations at ${\bf q}=\pi(1,1\pm 1/4)$ and ${\bf
  q}=\pi(1\pm1/4,1)$. Pairing
leads to bunching between opposite momenta 
${\bf k}=-{\bf k}'$ and s-wave pairing has been
observed near the BCS-BEC cross-over \cite{Ketterle}.  If dSF is
enhanced or suppressed by stripes, the bunching due to dSF
should be correlated correspondingly with stripe anti-bunching.

3D optical lattices do not have a van Hove singularity at the Fermi
surface at half filling as the 2D Hubbard model, and the 2D d-wave
symmetry $\Delta_k\propto(\cos k_x-\cos k_y)$ does not generalize to
3D. Thus we do not expect any significant dSF but by generalizing the
RMFT equations to 3D we find MI, AF and PS for the same reasons that
they appear in 2D.

In conclusion, t-J-U model RMFT calculations predict
phase separation and density discontinuities near half filling in 2D
and 3D optical lattices when $U\ga7.3t$ coinciding with a first
order transition in the double occupancy at half filling. 
Observation of phase separation would indicate that long range Coulomb
frustration is most likely the cause for spin and charge density waves
and stripes in cuprates. Contrarily, if no PS is
observed but instead a stripe phase near half filling in 2D and
probably also 3D optical lattices, then Coulomb
frustration is not responsible for stripes in cuprates. In either case
optical lattices not only emulate strongly correlated systems,
Hubbard type models and can determine the ground state phases such as
MI, AF, PS, gossamer and dSF but can even determine more subtle effects from
Coulomb frustration.

\vspace{-.4cm}


\begin{thebibliography}{99}
\vspace{-2.1cm}

\bibitem{Stoferle} T. St\"oferle et al.,
Phys. Rev. Lett. {\bf 96}, 030401 (2006).  
\bibitem{Folling} S. F\"olling et al., Nature {\bf 434}, 481 (2005).
\bibitem{Ketterle} J.K. Chin, D. E. Miller, Y. Liu, C. Stan, W. Setiawan, 
C. Sanner, K. Xu, W. Ketterle, Nature {\bf 443}, 961 (2006).

\bibitem{Spielman} I.B. Spielman, W.D. Phillips, and J.V. Porto,
Phys. Rev. Lett. {\bf 98}, 080404 (2007). 
\bibitem{Schneider} U. Schneider et al., Science {\bf 322}, 1520 (2008)

\bibitem{Kohl} M. K\"ohl, H. Moritz, T. St\"oferle, K. Gunter, T. Esslinger,
Phys. Rev. Lett. {\bf 94}, 080403 (2005).
\bibitem{Rom} T. Rom, Th. Best, D. van Oosten, U. Schneider, S. Foelling, 
B. Paredes, I. Bloch, Nature {\bf 444}, 733 (2006).

\bibitem{Edegger} See e.g., B. Edegger, V.N. Muthukumar, C. Gros,
 Advances in Physics {\bf 56}, 927 (2007), and Refs. therein. 


\bibitem{Laughlin}  R.B. Laughlin, arXiv:cond-mat/0209269; 
F.C. Zhang, Phys. Rev. Lett.  {\bf 90}, 207002 (2003);
J.Y. Gan, F.C. Zhang, Z.B. Su, Phys. Rev. B  {\bf 71}, 014508 (2005).
\bibitem{Klein} A. Klein and D. Jaksch, Phys. Rev. A {\bf 73}, 053613 (2006).
\bibitem{Yuan} Q. Yuan, F. Yuan, C. S. Ting, Phys. Rev. B  {\bf 72}, 054504 
(2005); F. Yuan, Q. Yuan, C.S. Ting, and T.K. Lee, arXiv:cond-mat/0409596.
\bibitem{Ogawa} T. Ogawa, K. Kanda, and T. Matsubara, Prog. Theor. Phys. 
{\bf 53}, 614 (1975).
\bibitem{Vollhardt} D. Vollhardt, Rev. Mod. Phys. {\bf 56}, 99 (1984).

\bibitem{HH} H. Heiselberg, Phys. Rev. A {\bf 74}, 033608 (2006); 
cond-mat/0802.0127.

\bibitem{Bloch} I. Bloch, J. Dalibard, and W. Zwerger,
Rev. Mod. Phys. {\bf 80}, 885 (2008); 
E. Altman, E. Demler, and M.D. Lukin, Phys. Rev. A {\bf 70}, 013603 (2004).
B.M. Andersen and G.M. Bruun, Phys. Rev. A {\bf 76}, 041602(R) (2007) 

\bibitem{Tranquada} J.M. Tranquada et al., Phys. Rev. B {\bf 54}, 7489 (1996).
\end{thebibliography}
\end{document}